# Broadband directional control of thermal emission

Jin Xu, Jyotirmoy Mandal, Aaswath P. Raman*

*Department of Materials Science and Engineering, University of California, Los Angeles, CA*

* Corresponding Author: Aaswath P. Raman (aaswath@ucla.edu)

**Abstract**

Controlling the directionality of emitted far-field thermal radiation is a fundamental challenge in contemporary photonics and materials research. While photonic strategies have enabled angular selectivity of thermal emission over narrow sets of bandwidths, thermal radiation is inherently a broadband phenomenon. We currently lack the ability to constrain emitted thermal radiation to arbitrary angular ranges over broad bandwidths. Here, we introduce and experimentally realize gradient epsilon-near-zero (ENZ) material structures that enable broad spectrum directional control of thermal emission by supporting leaky electromagnetic modes that couple to free space at fixed angles over a broad bandwidth. We demonstrate two emitter structures consisting of multiple semiconductor oxides in a photonic configuration that enable gradient ENZ behavior over long-wave infrared wavelengths. The structures exhibit high average emissivity ($> 0.6$ and $> 0.7$) in the p polarization between 7.7 and 11.5µm over an angular range of 70°-85°, and between 10.0 to 14.3µm over an angular range of 60°-75°, respectively. Outside these angular ranges, the emissivity dramatically drops to 0.4 at 50° and 40°. The structures' broadband thermal beaming capability enables strong radiative heat transfer only at particular angles and is experimentally verified through direct measurements of thermal emission. By decoupling conventional limitations on angular and spectral response, our approach opens new possibilities for radiative heat transfer in applications such as thermal camouflaging, solar heating, radiative cooling and waste heat recovery.

Thermally-generated light is a fundamental feature of nature[1–5]. The ubiquity of thermal emission makes its control of central importance to a broad range of technologies, ranging from energy conversion[6,7] to imaging[5,8] and sensing[9–11]. Thermal emitters are, in general, incoherent sources that lack directionality. However, high thermal emission at unwanted directions can lead to low efficiency and effectiveness for many thermal devices and applications[12]. Controlling the directional response of thermal emission has thus been investigated using a range of strategies[13], including surface plasmon polariton gratings and metasurfaces[14–17], phonon-polariton-based gratings[18], or photonic crystals[14]. While these past approaches have demonstrated anomalous angular control of thermal emission, their selectivity is over narrow bandwidths, with the angular response varying as a function of wavelength. Thermal radiation, however, is intrinsically a broadband phenomenon. The ability to constrain broad spectrum thermal emission to well-defined, arbitrary directions would be a fundamentally enabling capability[19], as it would facilitate high radiance and heat transfer in terms of bandwidth while restricting this heat transfer to specific angular ranges. Such a capability is of particular relevance to a range of sensing and energy applications including thermal imaging[20], thermophotovoltaics[21,22] and radiative cooling[23,24], where avoiding directional noise, and parasitic heat gain or loss, can result in significant gains in device efficiency.

Broadband angular selectivity of electromagnetic waves has, in general, proven to be a difficult task[13]. Prior work has enabled angular selectivity in the transmission of light over optical wavelengths by exploiting a Brewster-angle-enabled effect[25,26]. Broadband angular control of thermal emission from a bulk or nanostructured material remains challenging, with a recent theoretical proposal examining the use of model hyperbolic metamaterials where a broad spectrum Brewster angle could enable angular control of thermal emission[19]. To overcome this challenge, we propose and experimentally demonstrate a mechanism for the broadband directional control of thermal emission by using both photonic design as well as multiple materials whose intrinsic variation in permittivity enables epsilon-near-zero (ENZ) behavior[27,28].

We first introduce the theoretical concept of *gradient* ENZ materials whose permittivity crosses zero at a range of frequencies that vary across a spatial gradient. We show that a gradient ENZ film along the depth dimension supports broad spectrum leaky modes that couple to free-space propagating modes at fixed angles of incidence, thus functioning as a broadband directional thermal emitter whose angular range of 'thermal beaming' can be controlled by its total thickness. We then fabricate two deep-subwavelength photonic structures as realizations of the conceptual gradient ENZ film. Each structure consists of three polaritonic materials that collectively support gradient ENZ behavior at different and complementary parts of the long-wave infrared (LWIR) wavelength range. We experimentally characterize the spectral and angular response of both photonic structures and demonstrate broadband thermal beaming from 8.0 to 11.5µm over an angular range of 70° to 85° for the first structure, and from 10.5 to 14.3µm over an angular range of 60° to 75° for the second structure. We further perform direct emission measurements at fixed temperatures verifying the directional nature of thermal emission at different angles for each structure.

Consider a thermal emitter that presents low emissivity over all wavelength, $\varepsilon(\lambda,\theta) \sim 0$, at all polar angles of incidence $\theta$ except between a certain range of angles $\theta_{emit} \in [\theta_1, \theta_2]$ where it exhibits high emissivity, $\varepsilon(\lambda,\theta_{emit}) > 0.5$. Such a broadband directional thermal emitter, schematically shown in Figure 1, would then exhibit high radiance over the range of angles defined by $\theta_{emit}$, potentially equivalent to the high radiance emitter in the specified angular range, but at all other angles appear as reflective, and low emissivity, as a mirror. This is in contrast to conventional bulk materials as well as finite-size nanostructures, where light typically couples to propagating free space modes at a particular wave vectors *k,* and thus angles of incidence $\theta$ , that vary as a function of frequency $\omega$. Instead, we seek optical structures whose electromagnetic modes couple to propagating free-space modes at angles defined by $\theta_{emit}$ over a large bandwidth. Central to this broadband thermal beaming capability is the emitter's ability to present high emissivity over a broad range of wavelengths thereby enabling high total radiance, while maintaining angular selectivity and directionality throughout the thermal wavelength range of interest.

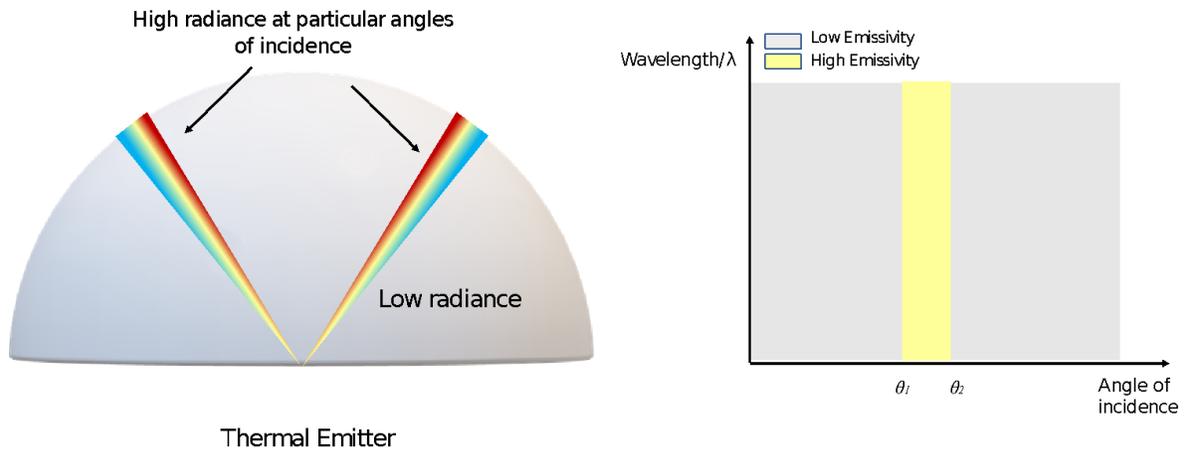

**Figure 1: Schematic capability of a broadband thermal directional emitter.** Such an emitter has low emissivity except at certain angles of incidence and emission, where it shows high emissivity across the entire thermal wavelength range of interest.

Thin, subwavelength films of materials exhibiting epsilon near zero (ENZ) behavior are known to support a leaky p-polarized electromagnetic mode near their ENZ region, termed a Berreman mode, which can couple to propagating free-space modes at a range of wave vectors[29–33]. Prior work has established this mode, and related ENZ behavior as narrow-band in nature, and has primarily been explored in this context at near infrared (NIR) wavelengths[30,34–37]. A related body of work, with earlier history, demonstrated the angular nature of absorption and emission in polaritonic materials over narrow ranges of long-wave infrared wavelengths[29,31,38]. This however raises an important constraint: ENZ behavior is typically narrow band in real materials due to its association with the wavelength range near a pole or resonance in the material's permittivity. How then can a broadband ENZ material be realized? Conceptually a material whose dielectric permittivity is near zero over a broad range of wavelengths can act as a broadband directional thermal emitter (see Supplementary Information Figure S1 for simulated angular response of a broadband directional thermal emitter made of an ideal ENZ thin film). While intriguing theoretically, in practice ENZ behavior is tied to material dispersion, and is thus inherently narrowband.

To overcome this limitation, we introduce *gradient ENZ materials* whose permittivity is defined by a material resonance pole that changes its frequency across a spatial gradient. More specifically, we consider 1D gradient ENZ films whose permittivity we express by modifying the standard model for the dielectric function of a polaritonic material with loss:

$$\varepsilon(\omega, z) = \varepsilon_{\infty}\left(1 + \frac{\omega_L^2 - \omega_T^2}{\omega_T^2 - (\omega - (\frac{d}{D})\omega_{range})^2 - i\gamma(\omega - (\frac{d}{D})\omega_{range})}\right), 0 \leq d \leq D$$

$$\omega_{range} = \omega_L - \omega_T$$

Here $\omega_T$ and $\omega_L$ are phonon frequencies correspond to out-of-phase atomic lattice vibrations with wave vectors aligned parallel (L, longitudinal) and perpendicular (T, transverse) to the incident field, $\varepsilon_{\infty}$ is the permittivity at infinite frequency, $\gamma$ is the damping rate, $d$ is the depth in the gradient ENZ film in $z$ direction and $D$ is the film's total thickness. At each depth, we thus have the permittivity crossing zero at different frequencies, establishing a range of ENZ frequencies that vary spatially. As the depth increases, this crossing point moves to smaller frequencies as shown in Figure 2(a). This gradient ENZ film, exhibits ENZ values across a wide dimensionless frequency range from 1.1 to 1.3, a range that can be tuned by altering the overall spatial gradient of the material pole response.

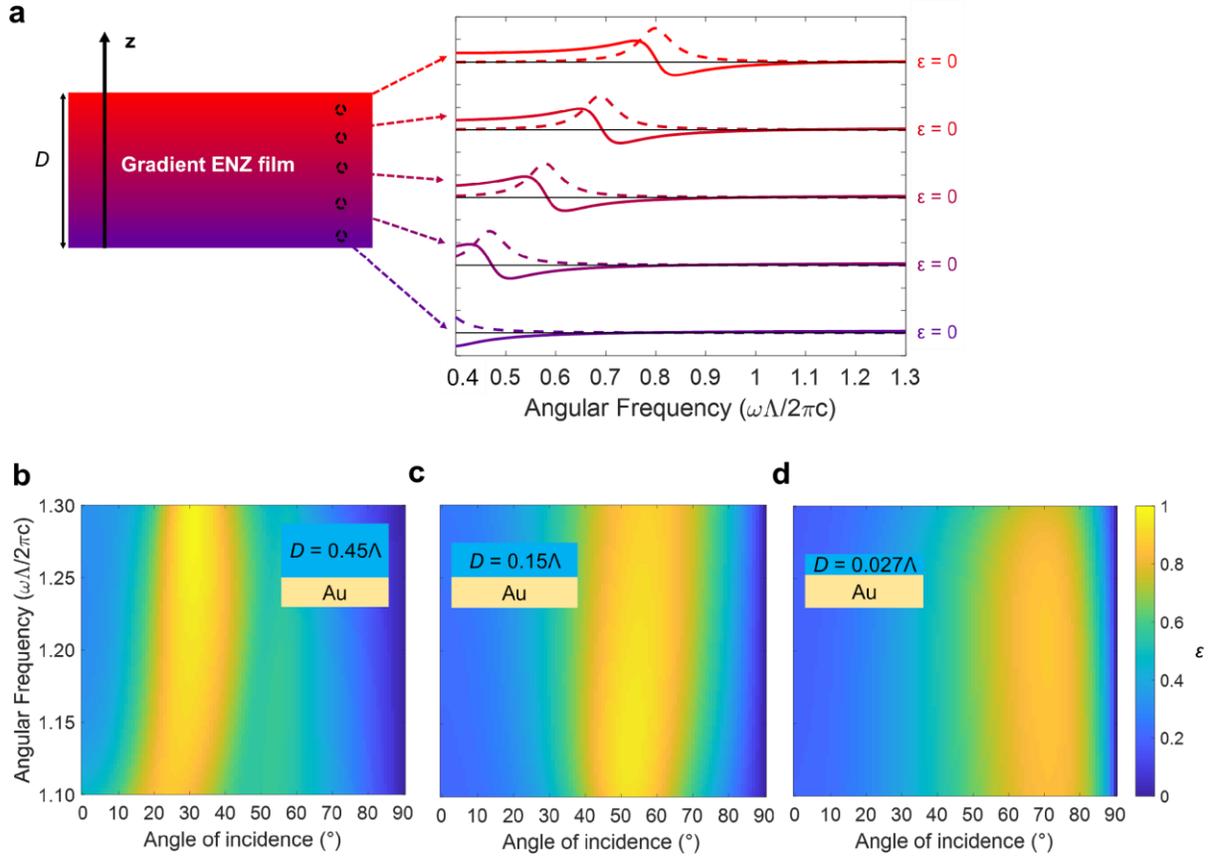

**Figure 2** (a) Schematic representation of the spatial gradient in the permittivity of the gradient ENZ film, whose dielectric permittivity is effectively near to zero across a wide dimensionless frequency range from 1.1 to 1.3, (b)-(d) The emissivity of the gradient ENZ film varying with the total thickness in p polarization as a function of angle of incidence and angular frequency. As the total thickness of the gradient ENZ materials increases from 0.027Λ to 0.15Λ and 0.45Λ respectively, the center of the high emissivity band moves from 70° to 50° and 30°.

To characterize the spectral and angular response of gradient ENZ films, we perform transfer-matrix simulations of three different total thicknesses of gradient ENZ film in the p-polarization (Figure 2(b)-(d)). The thinnest film (Figure 2(d)) displays the desired behavior schematically shown in Figure 1: it has high emissivity in the p polarization over a narrow range of angles, here centered at 70°, with low emissivity at other angles, in the dimensionless frequency range from 1.1 to 1.3 where the entire gradient ENZ film supports near zero permittivity at across its depth. As the total thickness of the gradient ENZ materials increases, the angular range of high emissivity moves towards normal incidence but consistently over the entire bandwidth (Figure 2(b)-(c)), indicating the ability of gradient ENZ films to 'beam' their thermal radiation towards normal incidence in the p-polarization. To better understand the underlying soure of this behavior, we compute the dispersion relations of the gradient ENZ films with different thicknesses (Supplementary Information Figure S5) in the p-polarization. We observe that gradient ENZ films support unique, *broadband* Berreman modes to the left of the light line which couple to free space corresponding to varying angles of incidence over the entire frequency range (1.1-1.3). In particular, we note that, as a function of thickness, the gradient ENZ film's supported mode is nearly parallel to the corresponding propagating free-space mode for a fixed angle of incidence over the entire frequency range, a unique and unexpected behavior in conventional materials as well as photonic structures. Moreover, we again observe that as the film's thickness increases, the mode moves to smaller *k* vectors, nearly parallel to the propagating free-space modes at near-normal angles of incidence.

We experimentally realize two gradient ENZ structures targeting the spectral range that corresponds to the blackbody spectral radiance of room-temperature range objects (300 K): long-wave infrared (LWIR) wavelengths. For both structures, we exploit the phonon polariton response of a range of semiconductor and metal oxides throughout this wavelength range that in turn yield ENZ frequencies. The first structure is composed of Silicon dioxide ($SiO_2$), Silicon monoxide (SiO) and Aluminum oxide ($Al_2O_3$) which present ENZ wavelengths of 8.0 µm, 8.7 µm, and 10.7 µm respectively. In Figure 3(a) we plot the permittivity of the three oxides based on tabulated values[39,40]. Layering them results in a gradient ENZ thin film because of their complementary resonances in the long-wave infrared part of the spectrum. Furthermore, due to their amorphous nature, their phonon resonances are relatively broadband, with their permittivity varying slowly as a function of wavelengths. While this is usually not desirable for conventional ENZ applications, it is beneficial for broadband scenarios, including the functionality we seek: the real part of the permittivity stays under unity for all three materials for a broad range of wavelengths. For the second structure, we use Tantalum Pentoxide $(TaO_5)$[41], Titanium oxide $(TiO_2)$[39], and Magnesium oxide $(MgO)$[42] which present ENZ wavelengths at 10.7 µm, 12.0 µm, and 13.5 µm respectively (Figure 3(b)). The thicknesses of both structures are optimized prior to fabrication through numerical simulations. Figure 3(c) and Figure 3(d) show Scanning Electron Microscope (SEM) images of the fabricated gradient ENZ structures and the layer thicknesses, with both structures deposited atop a base layer of aluminum, on a 4-inch Silicon wafer (see Supplementary Information for details).

The measured average emissivity in the p-polarization of structure 1 over its wavelength range of operation (7.7-11.5µm) is shown in Figure 3(e) as the red curve. As can be seen in the polar plot, within the angular range from 70° to 85°, it exhibits an average emissivity between 7.7 and 11.5µm in the p polarization > 0.6. Outside this angular range, average emissivity is less than 0.3, a 2:1 contrast. We note that this wavelength range accounts for ~35% of the radiative heat flux of an object at 300 K, and also overlaps the atmospheric window responsible for radiative cooling. In Figure 3(e), the blue curve is the measured average emissivity of the second gradient ENZ structure over its wavelength range of operation (10.0-14.3µm) for p polarization. The average p-polarized emissivity within the angular range between 60° and 75°, is above 0.7. Outside this angular range, emissivity drops dramatically to 0.4 at 40°.

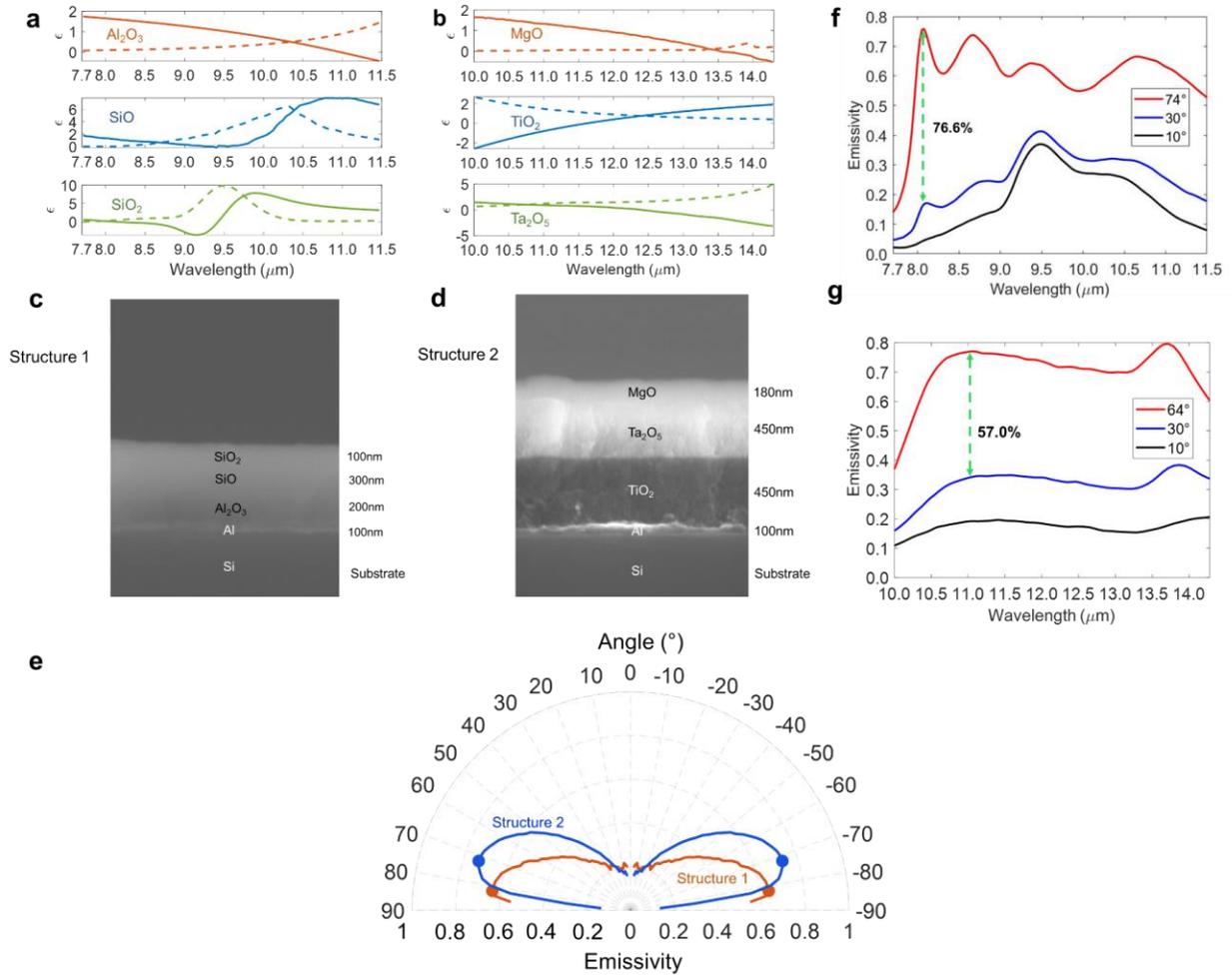

**Figure 3:** (a) and (b) The real (solid) and imaginary (dashed) parts of the permittivities of $SiO_2$, SiO and $Al_2O_3$ in the first multilayer structure and $Ta_2O_5$/$TiO_2$/MgO in second, each of which shows a slowly-varying permittivity that crosses zero at complementary wavelengths in the long-wave infrared part of the spectrum (c) and (d) SEM images of the experimentally fabricated multilayer photonic film structures. (e) Polar plot of the measured average emissivity over broad wavelength range (7.7-11.5µm and 10.0-14.3 µm respectively) varying with angle of incidence for p polarization in those two structures, with low emissivity at low angles of incidence and high emissivity at angles of incidence of 70° to 85° and 60° to 75° for structure 1 and structure 2 (f) and (g) Measured emissivity of the two photonic structures varying with wavelength at 3 different angles for p-polarization. Both the photonic structures exhibit a strong contrast between emissive and reflective states as a function of angle of incidence.

To better characterize the beaming effect, we plot in Figure 3(f) the spectral emissivity in the p polarization for angles of incidence of 10°, 30° and 74°. We observe a clear and consistent emissivity difference between different angles of incidence over the spectral range from 8.0 - 11.5 µm, with as much as a 76.6% contrast obtained near 8 µm. We show the spectral nature of the angular selectivity of structure 2 in Figure 3(g). Here as well we observe consistent, broadband (from 10.5 - 14.3 µm) contrast between the measured emissivity at angles of incidence of 10° and 30°, and 64°, for p-polarization. (The emissivity at all angles in both the p and s polarization is shown in Figure S3). We note that both fabricated structures present very low emissivity at all angles in the s-polarization (Figure S3) due to their deep subwavelength nature. Thus, while the polarization-averaged emissivity is lower than the measured p-polarization emissivity, the angular contrast and thermal beaming effect is maintained as it is the photonic structure's p polarization response that determines its angular response.

Furthermore, we observe that, as sought, structure 1 and the structure 2 have differing peak angles of emission, at 82° and 72° respectively, with the overall angular range of high emissivity also shifting from 70°-85° to 60°-75°. These results, in addition to demonstrating the core capability of broadband directional thermal emission, also highlight that through deliberate material choice and control of physical dimension, gradient ENZ films can deliver directional control at a range of angles of incidence with materials easily accessible today.

To model and understand the physical origin of the observed broadband thermal beaming effect enabled by the fabricated gradient ENZ structures, we numerically calculate the dispersion relation for the two photonic structures (see Supplementary Information). Figure S6 shows the dispersion relation of the two fabricated nanostructures, as well as alternative structures with differing thicknesses. For both structures, we observe frequency ranges above the light line which correspond to the ENZ wavelength ranges of $SiO_2$, SiO and $Al_2O_3$ in structure 1 and $Ta_2O_5$, $TiO_2$ and MgO in structure 2. Other frequency ranges lying to the right of the light line cannot be coupled to from free space. As the total thickness increases, the dispersion curve moves to the left, and thus will couple to modes from angles of incidence which are closer to normal incidence, agreeing well with the theoretical analysis of the conceptual gradient ENZ film (Figure S5), and indicating that angular response can be straightforwardly tuned to smaller angles of incidence by increasing the thickness of each layer.

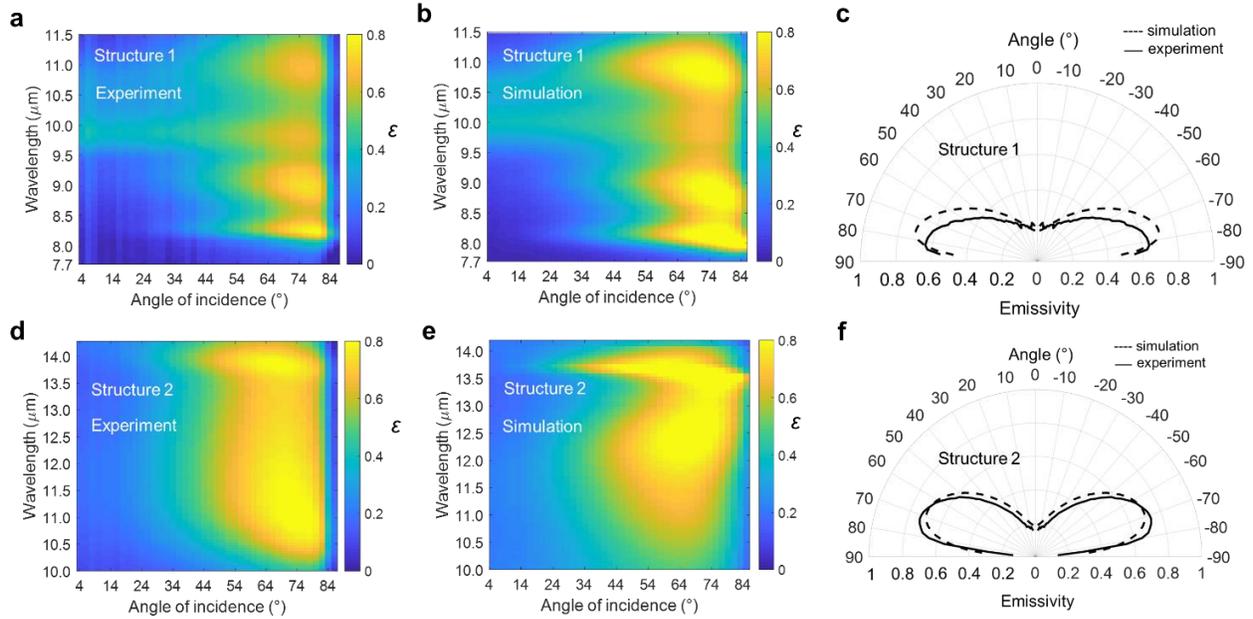

**Figure 4**. (a) and (d) Measured emissivity spectra varying with angle and wavelength of the two fabricated photonic structures in p-polarization, and (b) and (e) simulation results of the emissivity of the two structures for p-polarization using transfer matrix method, both show a broadband angularly selective behavior in emissivity and match well with the experiment measurement. (c) and (f) Average emissivity of the two structures from both measurement and simulation varying with angle of incidence for p polarization over wavelength ranges of 7.7-11.5 µm and 10.0-14.3µm respectively.

To validate our experimental results, we display in Figure 4 the measured as well as simulated emissivity spectra of both structures for the p-polarization varying with wavelength and angle of incidence. For structure 1, the spectrum exhibits strong peaks at 8.0 µm, 8.7 µm and 10.7 µm respectively, which correspond to the ENZ wavelengths of $SiO_2$, $SiO$ and $Al_2O_3$ respectively. The overall p-polarized emissivity throughout the 7.7-11.5 µm range, within the angular range from

70° to 85°, is above 0.6, as seen earlier in the polar plot of Figure 3. Transfer matrix simulations of the fabricated structure (Figure 4(b)) agree well with the experimental results and highlight clear angular regions of high and low emissivity that persist throughout the broad spectral range of operation. In Figure 4(c) we plot the average emissivity of structure 1 over the wavelength range of operation (7.7-11.5µm) from both measured and simulated data in polar coordinates. Though the measured emissivity is lower than expected from simulated data, the angles of operation and peak angle of incidence agree well. Similarly, we compare the measured and simulated emissivity spectra of structure 2 in Figure 4(d) and 4(e) for the p-polarization varying with wavelength and angle of incidence. Both measured and simulated spectrum exhibit continuous strong directional emission across the wavelength range from 10.5 µm to 11.3 µm owing to the slowly varying permittivity of $Ta_2O_5$, $TiO_2$ and MgO shown in Figure 3(b), which makes it a good gradient ENZ film. We again note a 10° shift in peak angle of incidence between structure 1 and 2, which is verified in numerical modeling. The average emissivity of structure 2 from both measured and simulated data in polar coordinates is shown in Figure 4(f) over the wavelength range of operation (10.0 - 14.3µm), showing good overall agreement.

A unique capability of the broadband nature of the directional thermal emitters we have demonstrated is their ability to emit a significant amount of heat radiatively at tailored angles of incidence. We experimentally probe this behavior through direct radiometric measurements of their directional thermal emission. Both structures, deposited on 4 inch Silicon wafers, are heated to 100°C for structure 1, and 110°C for structure 2, and emitted thermal radiation detected at fixed angles with a high-resolution uncooled microbolometer array (see Supplementary Information). Thermographs of both structures are shown in Figure 5(a) and (b) at a range of angles. We observe that at angles of incidence equaling 77.5° and 72.5° for structure 1 and 2 respectively, the wafer appears dramatically warmer compared with smaller angles of incidence (10° and 45°), even though the actual temperature of the wafer is unchanged.

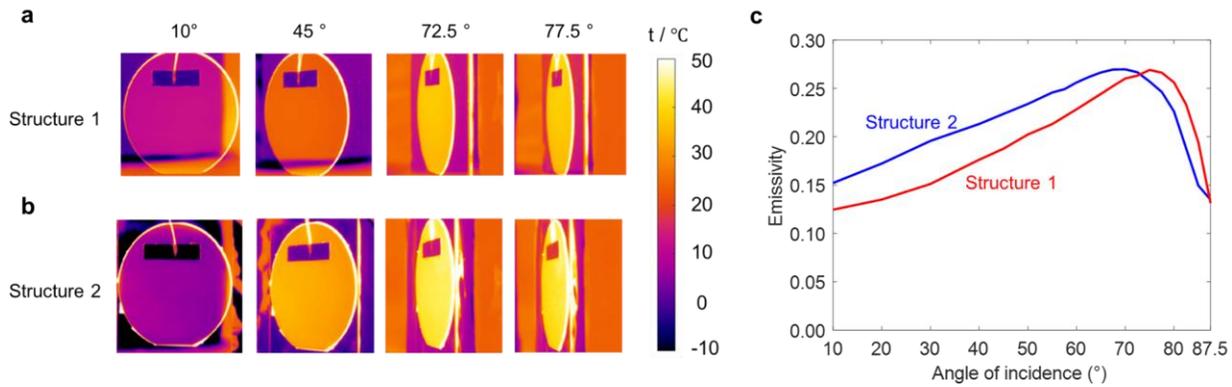

**Figure 5:** (a) and (b) Thermal images of the fabricated SiO$_2$/SiO/Al$_2$O$_3$ and Ta$_2$O$_5$/TiO$_2$/MgO multilayer photonic thin film structure. At the angle of incidence equals 77.5° and 72.5° for the two structures respectively, the wafer appears warmer compared with the angle of incidence is at 10° and 45° even though the actual temperature of the wafers maintains the same. (c) the calculated average emissivity for the two thermal emitter structures over the wavelength range between 7.5 µm and 13.5 µm by the temperature the thermal camera obtained at different angles of incidence.

Figure 5(c) shows the measured directional emissivity for the two structures over the wavelength range between 7.5 µm and 13.5 µm as obtained through radiometry (see Supplementary Information). Even though the measured emission by the detector is polarization averaged, the directionality of the thermal emission is clear. In agreement with the measurements of Figure 3 and Figure 4, we observe that the emissivity is high between 70° and 82° for structure 1, and between 60° and 80° for structure 2, with a peak emissivity value of 0.28 for both structures. The peak directional emissivity is smaller relative to the p polarization directional emissivity data shown in Figure 3 and Figure 4 because it is both polarization averaged, and because this measurement determines average emissivity over the entire 7.5µm - 13.5µm range. This range is 1.5 times larger than the designed wavelength range of operation of each structure (7.7µm - 11.5µm for structure 1 and 10.0µm - 14.3µm for structure 2). Despite the lower peak emissivity, we still note a nearly 2:1 emissivity contrast observed between normal and peak angles of incidence for both structures at 100°C and 110°C respectively, aligning with previously shown measurements. Thus, even though the enabled directional control is in the p polarization, it enables a marked, and anomalous, contrast in the polarization-averaged thermal emission as a function of angle since the emissivity is minimal at all angles in the s polarization over the relevant broadband spectrum (see Figure S3). This capability highlights immediate applications of the developed gradient ENZ structures for near-normal incidence heat signature control while maintaining significant radiative heat transfer at large angles.

We have thus proposed and experimentally demonstrated a mechanism, gradient ENZ materials, to enable broadband directional control of thermal emission and absorption. Our utilization of conventional oxides in fabricating our gradient ENZ films makes this approach immediately useful for large-area heat transfer applications for near-room temperature applications, since they are low-cost and amenable to well-established manufacturing processes. We emphasize that the temperature and wavelength ranges of operation can be tailored by material choice, including using indium tin oxide (ITO)[30,34] at near-infrared wavelengths for higher-temperature applications, as well as doped semiconductors for mid- and long-wave infrared applications. While our approach enables angular selectivity in the p polarization, higher total radiance can be achieved through simultaneous broadband angular selectivity in the s-polarization using effective mu-near-zero (MNZ) materials[43,44] which might be developed using a metamaterial approach. Alternatively, using gradient ENZ materials alone already delivers broad spectrum polarized light, which can be useful for a range of sensing and device applications. Our work points to new opportunities to leverage the unique capabilities of gradient ENZ materials and the optical modes they support for broadband applications. In particular, we believe that these effects may have near-term impact on heat transfer and energy applications, including directionality control for radiative cooling[24], thermophotovoltaics[45], and nascent approaches to waste heat recovery from radiative sources[46]. More broadly, these results highlight that by decoupling angular and spectral response in photonic nanostructures, new capabilities can emerge for how materials emit and absorb light.

## Acknowledgements

This work was supported by the Sloan Research Fellowship in Physics from the Alfred P. Sloan Foundation. Jyotirmoy Mandal was supported by Schmidt Science Fellows, in partnership with the Rhodes Trust. We thank Patricia McNeil and Bruce Dunn for experimental assistance.

**Supplementary Information**

**1.     Broadband angular selectivity with an ideal ENZ film**

We first consider an idealized ENZ thin film on a substrate of gold, with constant Re($\varepsilon$) = 0.01 and Imag($\varepsilon$) = 0.01i over a relevant thermal wavelength spectrum, from 8 to 12 μm corresponding to the peak regions of the blackbody spectral radiance of a 300 K object. We plot in Figure S1 (a) a transfer-matrix calculation of the emissivity and absorptivity of such a film of thickness $d$ = 500 nm for p-polarization. The film displays the desired behavior schematically shown in Figure 1: it has high emissivity in the p polarization over a narrow range of angles centered around 10°, with low emissivity at other angles, for all wavelengths of interest. For s polarization, the film has near-zero emissivity at all wavelengths. Thus, the angular nature of its polarization-averaged emissivity is defined by the p polarization response.

Directional thermal beaming to different angular ranges can be achieved by changing the thickness of the ENZ film. As shown in Figure S1 (b) and (c) at 50 nm and 5 nm thickness, the angular band of high emissivity is centered at 30° and 70° respectively. To capture the relationship between thickness and angular response, Figure S1 (d) shows the emissivity spectrum of the ideal ENZ thin film as its thickness varies from 0 nm to 500 nm at a wavelength of 8 μm for the p-polarization. As the film gets thicker, the emissivity peak moves from large angles of incidence to small angles.

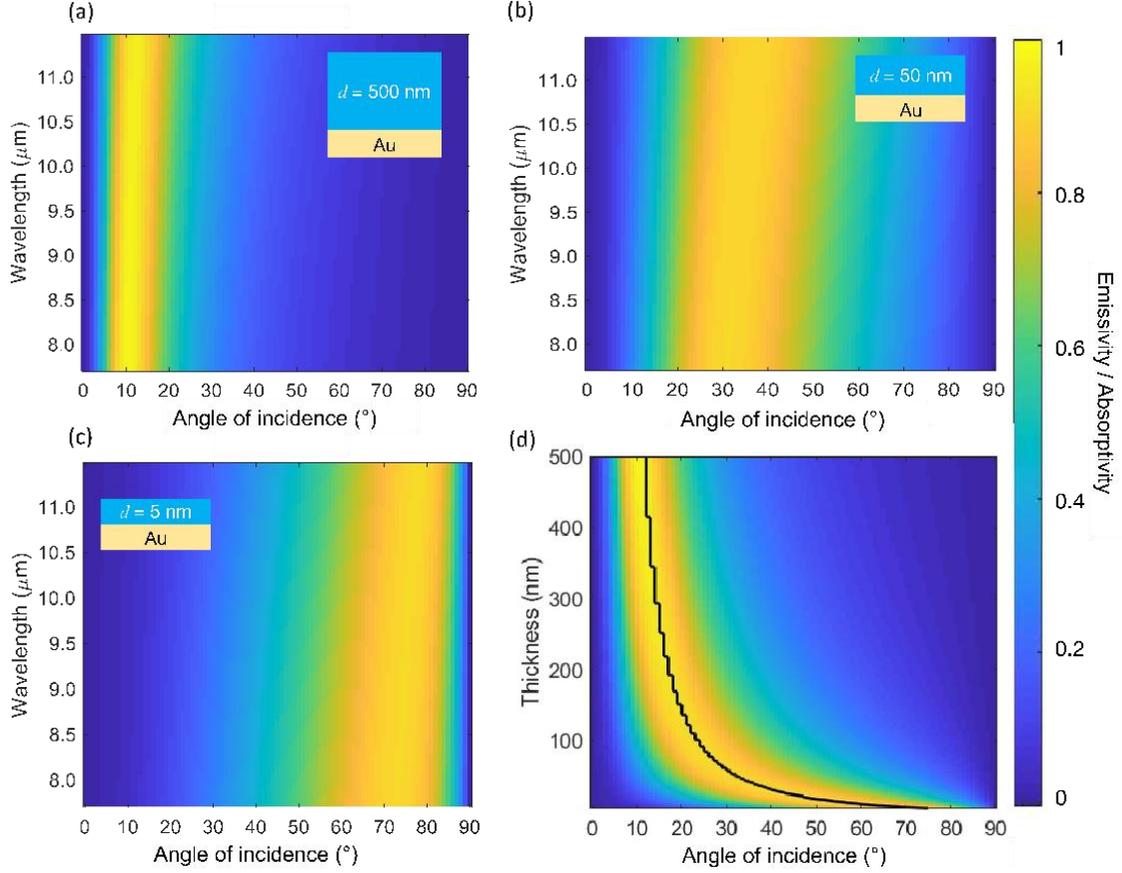

**Figure S1:** (a)-(c) Emissivity spectrum of an ideal ENZ thin film for p-polarization. The thickness of each ideal ENZ thin film is $d = 500$ nm, $d = 50$ nm, $d = 5$ nm, respectively. (d) Emissivity spectrum of the ideal ENZ thin film as the thickness varies from 0nm to 500nm at 8µm for p-polarization. As the thin film gets thicker, the emission peak moves from large to small angles.

## 2. Impact of permittivity values on broadband angular selectivity

Epsilon near zero (ENZ) materials can exhibit high absorption and emission of electromagnetic waves near their ENZ frequencies. For propagating modes in ENZ films, the absorption $A$ can be defined as[31]:

$$A(r,\omega) = 0.5\omega Im[\varepsilon(\omega)]|E(r,\omega)|^2 \quad (S1)$$

Due to the continuity of $\varepsilon E_\perp$ in the p polarization, the electric field $E_\perp$ is both enhanced and strongly confined in the ENZ films:

$$E(r,\omega) = \frac{E_0(r,\omega)}{Re(\varepsilon(\omega))} \tag{S2}$$

Here $E_0$ is the electrical field free space. Substituting Eq. S2 to Eq. S1, the absorption is related to the Im[ε] and Re[ε] as

$$A(r,\omega) = 0.5\omega Im[\varepsilon(\omega)]|\frac{E_0(r,\omega)}{Re[\varepsilon(\omega)]}|^2 \tag{S3}$$

From Eq. S3 we can see that when Re[ε] approaches zero, the absorption can be strongly enhanced. The absorption value however is also strongly influenced by Im[ε]. The strength of Im[ε] when Re[ε] approaches zero influences the overall angular width of absorption one will obtain at a particular frequency in a thin ENZ film. The larger Im[ε] is, the larger the angular width will be. This in turn can place limits on this approach to angular selectivity and broadband thermal beaming. To explore this relationship, we conduct a series of simulations for thin films which have a range of real and imaginary permittivities. We determine the optimal thickness for these thin films to support high absorption and emission at an angle of incidence near normal, with the results shown in Figure S2.

We can see that, given the same Im[ε], the larger Re[ε] is, the harder it is for the thin film to have an angle of incidence with high emissivity only near normal incidence. For the same Re[ε], for modes which can be coupled into the film from free space, the high emissivity angle of incidence is more likely to be constrained to near normal incidence when Im[ε] is smaller. This is driven in large part by the fact that, due to their dispersion relation, ENZ films that couple to free-space modes near-normal incidence need to have a larger thickness. If Im[ε] is substantially greater than zero, this can cause absorption to occur over a broader range of $k$ values, and thus angles of incidence, due to the larger thickness of the film.

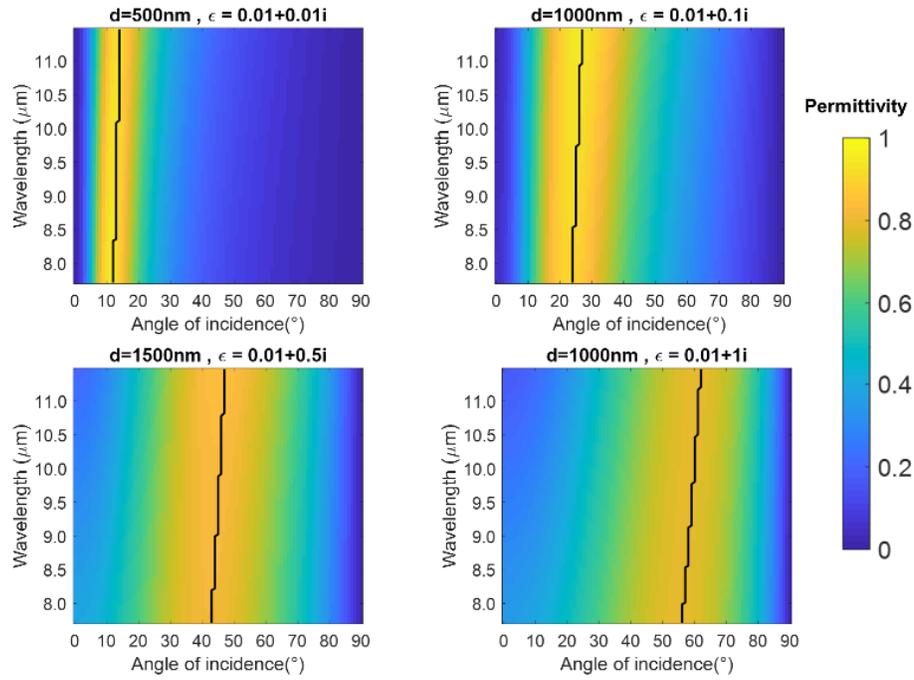
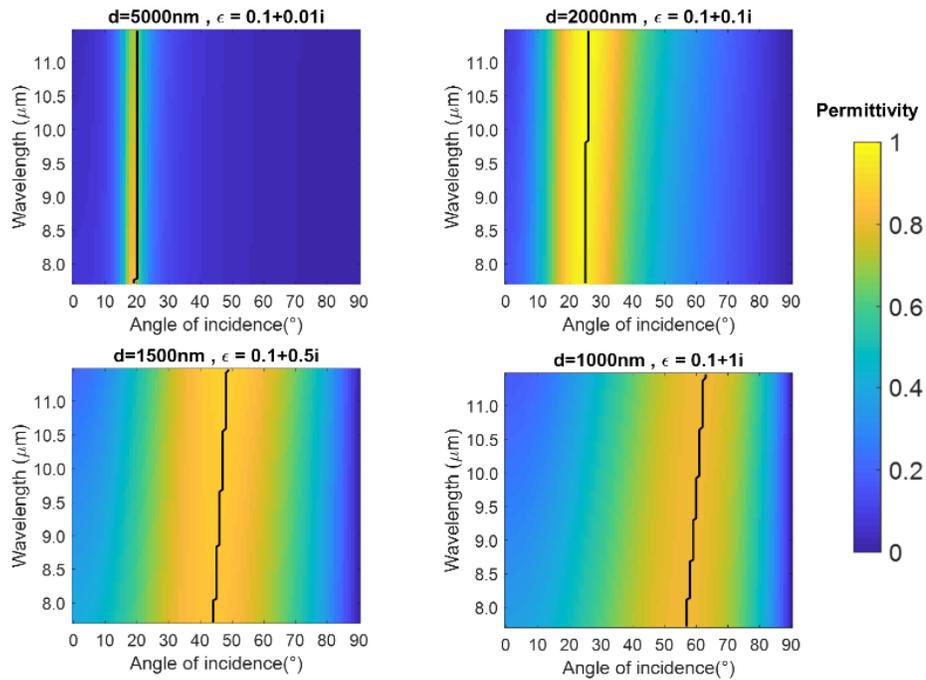

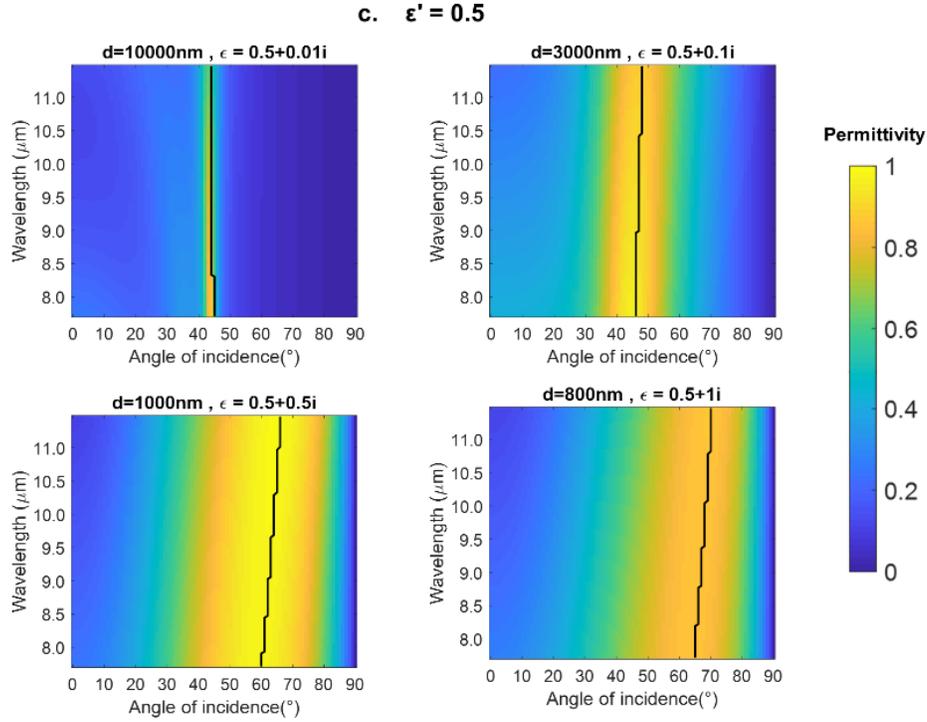

**Figure S2:** Emissivity spectra of thin films with varying real and imaginary permittivities near zero for p-polarization, shown as a function of angle of incidence over a fixed wavelength range. Results are grouped by real permittivities ($\varepsilon'$): (a) $\varepsilon' = 0.01$, (b) $\varepsilon' = 0.1$, and (c) $\varepsilon' = 0.5$.

## 3. Experimental methods

The two ENZ photonic structures (whose SEM images are shown in Figure 3(c) and (d)) were both coated on the 4-inch prime silicon wafers via electrical beam evaporation method using a CHA industries MARK 40 evaporator. The start evaporating pressure is at 7.7E-7 torr. First, a 5 nm adhesion layer of Ti was evaporated with a deposition rate of 0.7Å/s, followed by a 100 nm Al layer (deposition rate of 2.0 Å/s), 200nm of $Al_2O_3$ (deposition rate of 2.0 Å/s), 300nm of SiO (deposition rate of 2.5 Å/s) and 100nm of $SiO_2$ (deposition rate of 2.0 Å/s) for the first multilayer thin film structure and 180 nm of MgO (deposition rate of 1.3 Å/s), 450 nm of $TiO_2$ (deposition rate of 1.8 Å/s) and 450nm of $Ta_2O_5$ (deposition rate of 2.0 Å/s) for the second structure.

*Characterization*

By Kirchhoff's law, the absorptivity of an object is equal to its emissivity or each frequency and angle of incidence when in thermal equilibrium. Since our structures were deposited on an optically thick metal layer, the transmissivity ($T$) is equal to 0. Thus, to infer emissivity ($\varepsilon$), we measure reflectivity ($R$) as a function of polar angle of incidence, and plot it as $\varepsilon = 1 - R$.

Deposited samples were characterized by a Bruker Invenio-R Fourier-transform infrared spectrometer (FTIR) with a Harrick Seagull variable angle reflection accessory. An inserted wire-grid polarizer is used to selectively transmit linearly polarized light (p and s polarization namely) to the sample for measurements. Our measurements are made by establishing background reflectance from a well-calibrated reference mirror first for each measurement. The reflectivity spectra were measured for different incidence of angles from 4° to 84° at 2° increments.

## 4. Additional experimental data

*S polarization experimental measurements*

Figure S3 shows experimental data as well as simulation results for the emissivity of the two structures in the s polarization. We observe that in the s polarization both structures show low emissivity regardless of the angle of incidence. Thus it is the p polarization's angularly selective behavior which defines the angular contrast when measuring or making use of the polarization averaged emissivity.

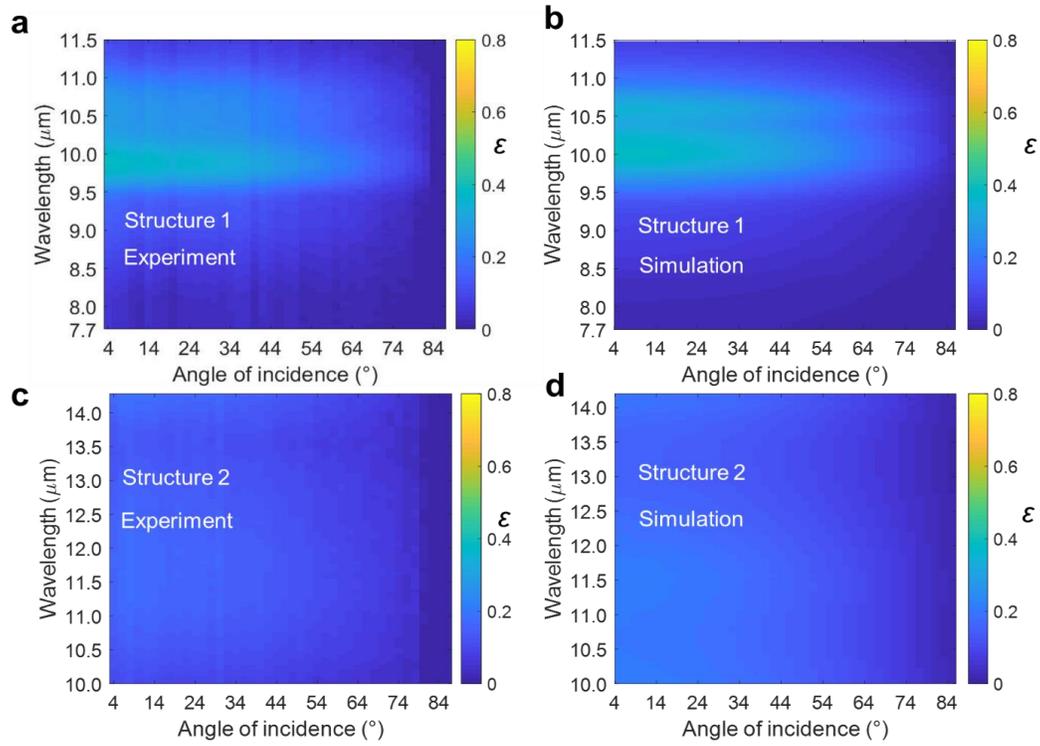

**Figure S3**. (a) and (c) Measured emissivity spectra varying with angle and wavelength of the two fabricated photonic structures for s-polarization, and (b) and (d) simulation results of emissivity in s-polarization using transfer matrix method show low emissivity regardless the angle of incidence and match well with the measure data

*Experimental data for both structures and polarizations*

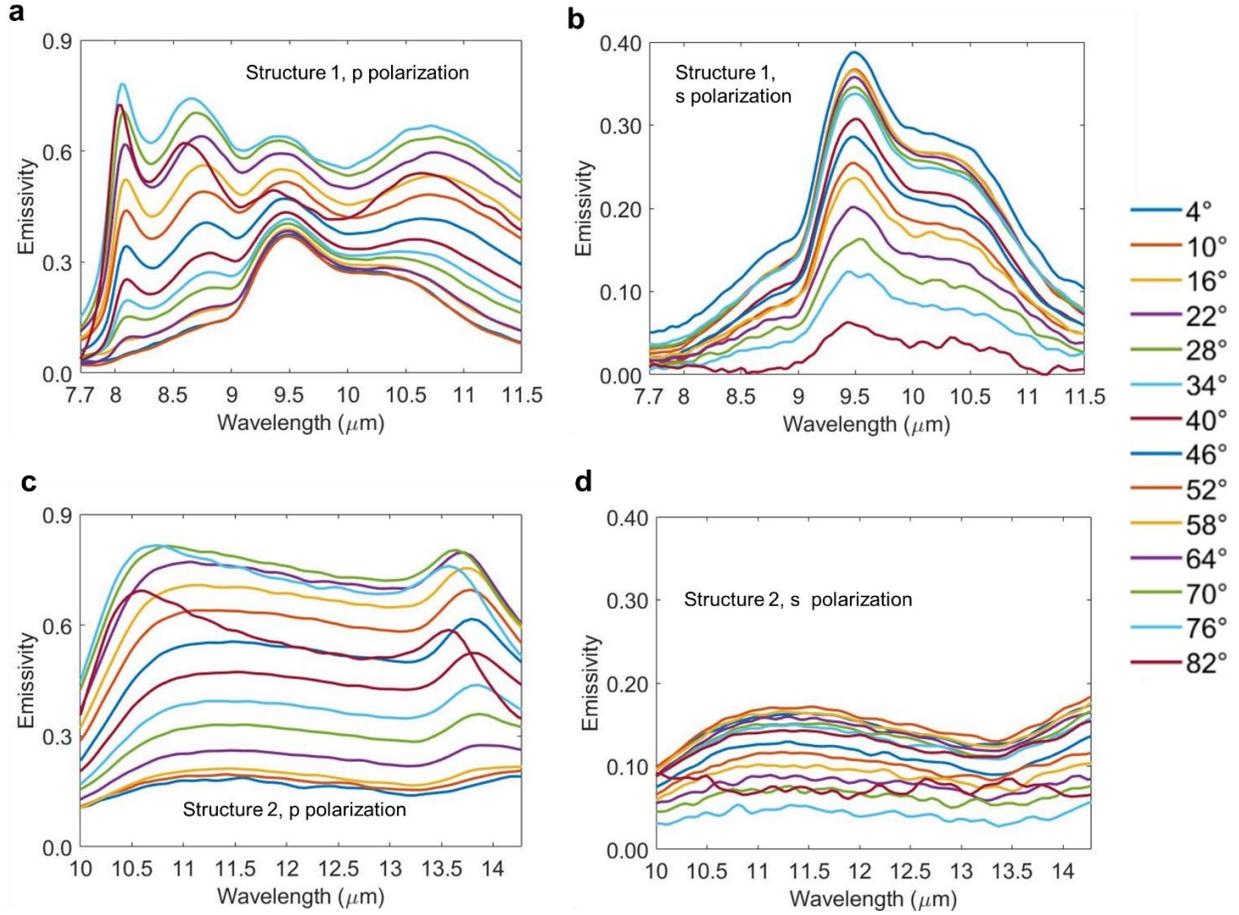

**Figure S4**. (a) and (b) Measured emissivity varying with angle and wavelength of structure 1 in p and s-polarization respectively, and (c) and (d) Measured emissivity varying with angle and wavelength of structure 2 in p and s-polarization respectively

## 4.  Theoretical model of photonic structures' electromagnetic modes

We calculate the photonic film's dispersion relation by finding the wavevector which presents minimum reflectivity for a given frequency $\omega$ and perform this search iteratively throughout our frequency range of interest[33]. The reflectivity is calculated by the transfer matrix method for p polarization. The transfer matrix for the *j*th film in a multilayer film in the p polarization is given by

$$M_j = \begin{pmatrix} \cos(k_j a_j) & i\sin(k_j a_j)\frac{\varepsilon_j \omega}{ck_j} \\ i\sin(k_j a_j)\frac{ck_j}{\varepsilon_j \omega} & \cos(k_j a_j) \end{pmatrix} \quad (S4)$$

where $k_j = \sqrt{\varepsilon_j \frac{\omega^2}{c^2} - k_x^2}$, $\varepsilon_j$ is the permittivity, $\omega$ is the free space frequency of the light, $c$ is the speed of light and $a_j$ is the thickness of the *j*th layer. The transfer matrix for the whole film consisting of *N* layers is then given by the product of the respective transfer matrices for the individual layers

$$M = \prod_{i=1}^{N} M_j = \begin{pmatrix} m_{11} & m_{12} \\ m_{21} & m_{22} \end{pmatrix} \quad (S5)$$

The reflectivity *R* of the overall film is then calculated as:

$$R = r^2 = \left(\frac{m_{21}}{m_{11}}\right)^2 \quad (S6)$$

Figure S5 shows the dispersion relation of the gradient ENZ thin film described in Figure 2. As we vary the total thicknesses of the film, from the red curve (0.027 $\Lambda$) to green (0.15$\Lambda$) and blue curve (0. 45$\Lambda$), where $\Lambda$ is the dimensionless average wavelength, the broadband Berreman mode moves towards normal incidence, which is corresponding to the results in the Figure 2(b)-(d).

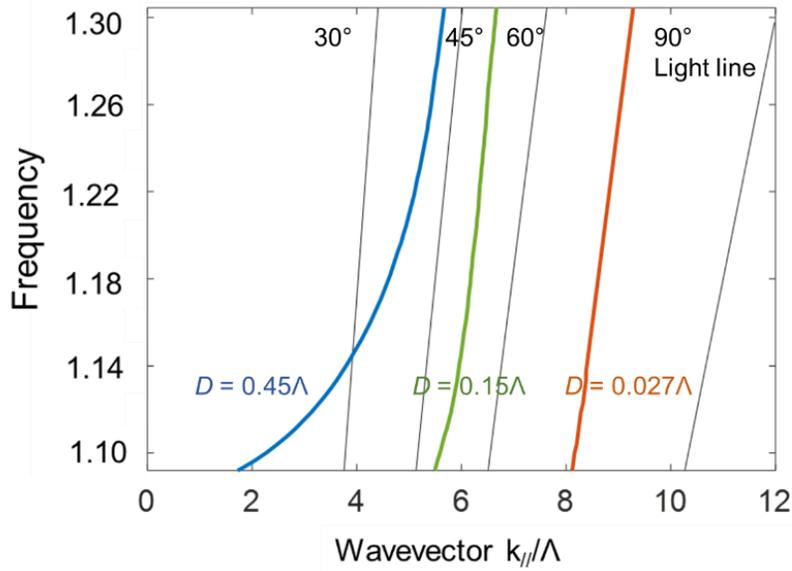

**Figure S5:** Dispersion relation of the gradient ENZ thin films. The total thicknesses, for the red, green and blue curves, increases from 0.027 Λ to 0.15Λ and 0.45Λ respectively, where Λ is the dimensionless average wavelength. The broadband Berreman mode moves towards normal incidence as the total thickness of the gradient ENZ film increases, which agrees to the results in Figure 2(b)-(d).

Figure S6 shows the calculated dispersion relation of the two fabricated gradient ENZ structures, as well as structures of alternative thicknesses with the same materials. For structure 1 ($SiO_2$/SiO/$Al_2O_3$), we observe three frequency ranges above the light line which correspond to the three ENZ wavelength ranges of $SiO_2$, SiO and $Al_2O_3$. Other frequency ranges lying to the right of the light line cannot be coupled to from free space. Importantly, we see that as the total thickness increases, the supported electromagnetic modes move to the left, and thus will couple to propagating free space modes that are closer to normal incidence, agreeing well with the theoretical analysis of the gradient ENZ structure, and indicating that angular response can be tuned by the thickness of each layer's material. Similar behavior is observed in the dispersion relation of structure 2 ($Ta_2O_5$/$TiO_2$/MgO). Owing to the gradual spectral variation in the emissivities of the three materials in structure 2, we observe a near-continuous broadband Berreman mode located to the left of the light line. This is in good agreement with what we see in both the experimental measurement and simulation in Figure 4(d) and (e).

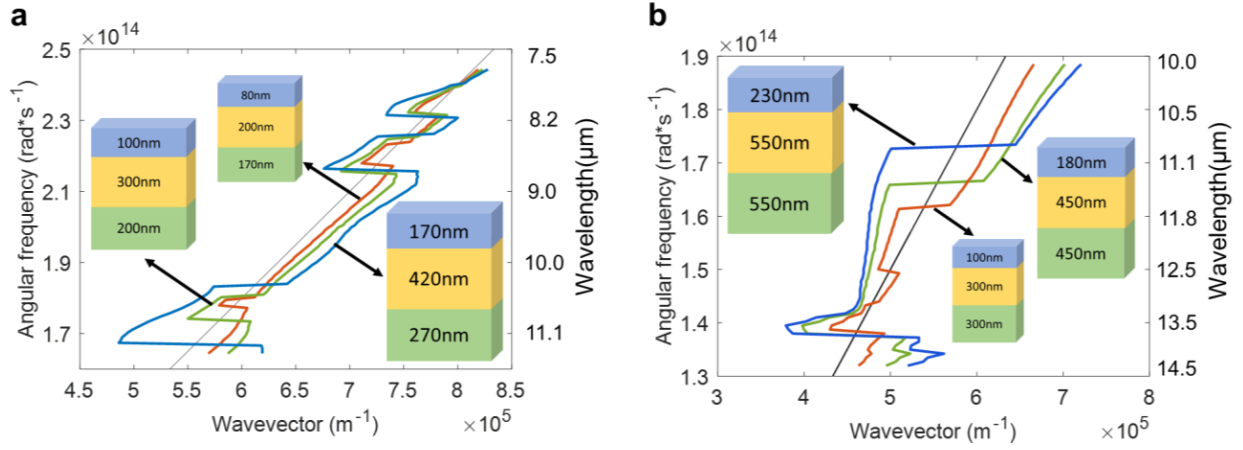

**Figure S6:** Dispersion relation of (a) gradient ENZ structure 1 ($SiO_2/SiO/Al_2O_3$) and (b) structure 2 ($Ta_2O_5/TiO_2/MgO$).

### 6. Radiometric measurement of directional emissivity

The emissivity $\varepsilon$ of the two structures were radiometrically measured using a thermal camera with an uncooled microbolometer array (FLIR Boson 640). The set up for the thermography is shown in Figure S6. The wafer were mounted on a metal block heated by a heater to a set temperature ($T_{emitter}$~100°C for structure 1 and ~110°C for structure 2) and placed on the center of an arc on which the microbolometer array was set facing the emitter at different angles of incidence (Figure S7). The emitter also had an aluminum tape with known angular emissivity placed on it as a reference. Temperatures of the emitter, tape and the air in the laboratory setting were separately measured using thermocouples. Since the apparent temperature of an emitter is also a function of the background temperature $T_{bckgd}$ partially reflected by the emitter (S7), a metal plate coated with a high emittance coating ($\varepsilon$ ~0.97) in thermal contact with dry ice was used as the cold background.

For the measurements, the camera was placed at an angle $\theta$ from the surface normal to the emitter, and the plate providing the cold background was placed at $-\theta$. The apparent temperature $T_{rad}$ of the emitter near the aluminium tape, the aluminium tape itself, and the cold background ($T_{bckgd}$, when in direct view of the camera) was then measured using the thermal camera assuming an emittance of 1. When not in direct view, the temperature of the cold background was calculated using S7, but using the apparent temperature and known emittance of the aluminum tape. With both $T_{rad}$, $T_{emiter}$ and $T_{bckgd}$ thus known by calculation or measurement, S7 was used to calculate the emissivity of the emitter.

$$T_{rad}^4 = \varepsilon T_{emitter}^4 + (1-\varepsilon)T_{bckgd}^4 \tag{S7}$$

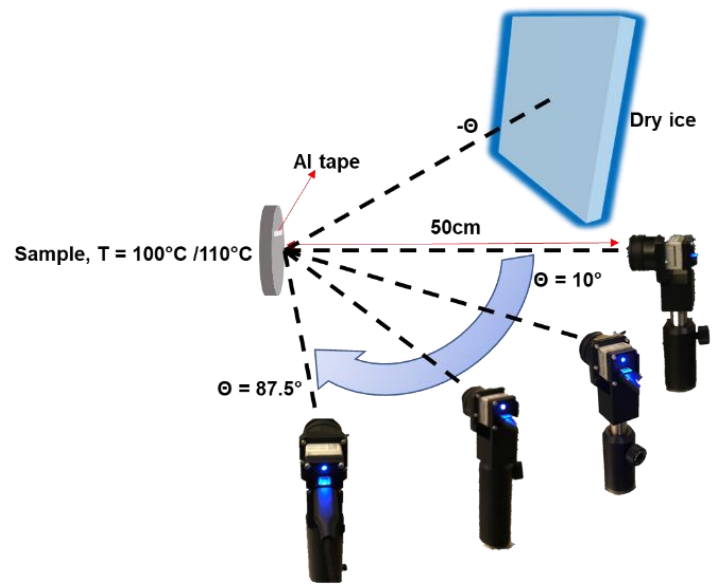

**Figure S7:** Schematic demonstration of the thermography set up

In Figure S8 we show an image of five identical wafers of our structure 1 (SiO$_2$/SiO/Al$_2$O$_3$) lying atop a low-emissivity metal support outdoors. We observe a strong angular contrast between reflective and emissive modes. The closer the wafer is to the camera the smaller the angle of incidence is, and thus these wafers appear colder to the thermal camera, and appear similar to the background support, a low-emissivity metal. The wafer which is the farthest away from the camera has the largest angle of incidence relative to the camera, and thus shows a warmer temperature due to its higher angular emissivity, and contrasts strongly with the low-emissivity metal support, which reflects the sky and thus appears cold. This capability highlights immediate applications of the developed material for near-normal incidence heat signature control while maintaining significant radiative heat transfer at large angles.

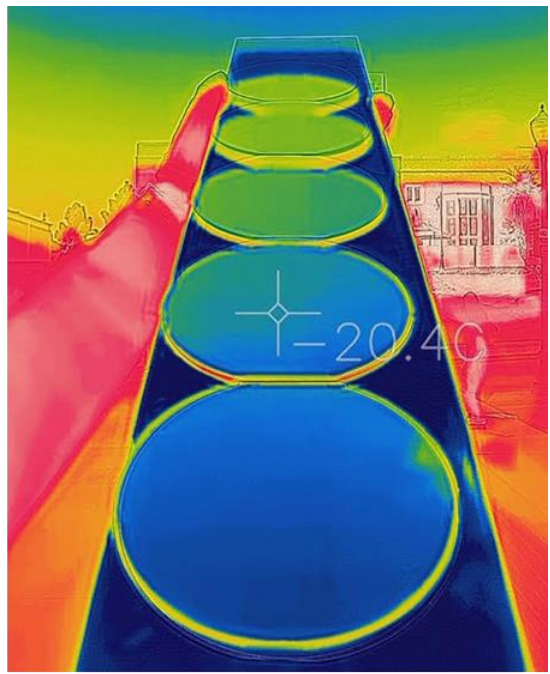

**Figure S8:** Thermal image of 5 identical fabricated wafers of structure 1 in an outdoor setting. The further the wafer is the larger the angle of incidence to the thermal camera is, which gives the farthest wafer the highest emissivity. The metal support, which is low emissivity and reflecting the sky, looks cold to the camera regardless of the angle of view.